\def\etal{{\frenchspacing\it et al.}}

\def\beq#1{\begin{equation}\label{#1}}
\def\eeq{\end{equation}}
\def\beqa#1{\begin{eqnarray}\label{#1}}
\def\eeqa{\end{eqnarray}}

\def\fun#1#2{\lower3.6pt\vbox{\baselineskip0pt\lineskip.9pt
        \ialign{$\mathsurround=0pt#1\hfill##\hfil$\crcr#2\crcr\sim\crcr}}}
        
\def\figsize{11.5cm}

\def\xi{{{\bf x}^b}}

\documentclass[twocolumn,aps,showpacs,showkeys,nofootinbib]{revtex4}
\usepackage{epsfig}

\newcommand{\be}{\begin{equation}}
\newcommand{\ee}{\end{equation}}
\newcommand{\ba}{\begin{eqnarray}}
\newcommand{\ea}{\end{eqnarray}}

\begin{document}
\input{epsf.sty}

\title{Observational Constraints on Dark Energy and Cosmic Curvature}
\author{Yun~Wang$^{1}$, and Pia~Mukherjee$^{2}$}
\address{$^1$~Homer L. Dodge Department of Physics \& Astronomy, Univ. of Oklahoma,
                 440 W Brooks St., Norman, OK 73019;
                 email: wang@nhn.ou.edu}
\address{$^2$~Department of Physics \& Astronomy, Univ. of Sussex, 
                 Falmer, Brighton, BN1 9QH, UK;
                 email: p.mukherjee@sussex.ac.uk}
                 \today

\begin{abstract}
Current observational bounds on dark energy depend on 
our assumptions about the curvature of the universe.
We present a simple and efficient method for incorporating constraints 
from Cosmic Microwave Background (CMB) anisotropy data, and 
use it to derive constraints on cosmic curvature and dark energy
density as a free function of cosmic time using current 
CMB, Type Ia supernova (SN Ia), and baryon acoustic oscillation (BAO) data.

We show that there are {\it two} CMB shift parameters,
$R\equiv \sqrt{\Omega_m H_0^2} \,r(z_{CMB})$ 
(the scaled distance to recombination)
and $l_a\equiv \pi r(z_{CMB})/r_s(z_{CMB})$
(the angular scale of the sound horizon at recombination),
with measured values that are nearly uncorrelated
with each other. 
Allowing nonzero cosmic curvature,
the three-year WMAP data give 
$R =1.71 \pm 0.03$, $l_a =302.5 \pm 1.2$, and 
$\Omega_b h^2 = 0.02173 \pm 0.00082$,
independent of the dark energy model.
The corresponding bounds for a flat universe
are $R =1.70 \pm 0.03$, $l_a =302.2 \pm 1.2$, and 
$\Omega_b h^2 = 0.022 \pm 0.00082$.
We give the covariance matrix of ($R$, $l_a$, $\Omega_b h^2$)
from the three-year WMAP data. 
We find that ($R$, $l_a$, $\Omega_b h^2$) 
provide an efficient and intuitive summary of CMB data 
as far as dark energy constraints are concerned.

Assuming the HST prior of $H_0=72\pm 8\,$(km/s)Mpc$^{-1}$,
using 182 SNe Ia (from the HST/GOODS program, the first year 
Supernova Legacy Survey, and nearby SN Ia surveys), 
($R$, $l_a$, $\Omega_b h^2$) from WMAP three year data,
and SDSS measurement of the baryon acoustic oscillation scale,
we find that dark energy density is consistent with a 
constant in cosmic time, with marginal deviations from a 
cosmological constant that may reflect current systematic
uncertainties or true evolution in dark energy. 
A flat universe is allowed by current data: 
$\Omega_k=-0.006_{-0.012}^{+0.013}$$_{-0.025}^{+0.025}$
for assuming that the dark energy equation of state $w_X(z)$ 
is constant, and
$\Omega_k=-0.002_{-0.018}^{+0.018}$$_{-0.032}^{+0.041}$
for $w_X(z)=w_0+w_a(1-a)$ (68\% and 95\% confidence levels).
The bounds on cosmic curvature are 
less stringent if dark energy density is allowed to be a free
function of cosmic time, and are also dependent on the assumption
about the early time property of dark energy. We demonstrate
this by studying two examples.
Significant improvement in dark energy and cosmic curvature constraints
is expected as a result of future dark energy and CMB experiments.

\end{abstract}

\pacs{98.80.Es,98.80.-k,98.80.Jk}

\keywords{Cosmology}

\maketitle


\section{Introduction}

The unknown cause for the observed cosmic acceleration \cite{Riess98,Perl99},
dubbed ``dark energy'', 
remains the most compelling mystery in cosmology today.
Dark energy could be an unknown energy component
\cite{Freese87,Linde87,Peebles88,Wett88,Frieman95,Caldwell98},
or a modification of general relativity 
\citep{SH98,Parker99,Boisseau00,DGP00,Mersini01,Freese02}.
\cite{Pad} and \cite{Peebles03} contain reviews of many models.
Dark energy model-building is a very active research area.
For recent dark energy models,
see for example, \cite{Carroll04,OW04,Cardone05,Kolb05,Caldwell06,KO06,DeFelice07,Koi07,Ng07}.
Current observational data continue to be consistent with dark energy being a cosmological
constant, but the evidence for a cosmological constant is not
conclusive and more exotic possibilities are still allowed (see, for example,
\cite{WangTegmark04,WangTegmark05,Alam05,Daly05,Jassal05a,Jassal05b,Barger06,Dick06,Huterer06,Jassal06,Liddle06,Nesseris06,Schimd06,Sumu06,Wilson06,Xia06,Alam07,Davis07,Wei07,Zhang07,Zun07}).

While the universe is completely consistent with being flat under a $\Lambda$CDM 
hypothesis, it is important to note that the observational bounds on dark energy 
and the curvature of the universe are closely related. 
Cosmic Microwave Background (CMB) anisotropy data provide the most
stringent constraints on cosmic curvature $\Omega_k$.
Assuming that dark energy is a cosmological constant, the 
three-year WMAP data give $\Omega_k = -0.15 \pm 0.11$, and this improves 
dramatically to $\Omega_k = -0.005 \pm 0.006$ with the addition of galaxy 
survey data from the SDSS \cite{sdss} (2dF data \cite{2df} also give 
a similar improvement) \cite{Spergel06}. 
The effect of allowing non-zero curvature on constraining some dark energy models
has been studied by \cite{Polar05,Franca06,IT06,Ichi06,Clarkson07,Gong07,Kazu07,Zhao07,Wright07}.

In this paper, we present a simple and efficient method for incorporating 
constraints from the CMB data into
an analysis with other cosmological data  
in constraining dark energy without assuming a flat universe.
Uisng this method, we derive constraints on dark energy and cosmic curvature
using CMB, type Ia supernova (SN Ia) and galaxy 
survey data.

We describe our method in Sec.II, present our results in Sec.III,
and conclude in Sec.IV.

\section{Method}

The comoving distance from the observer to redshift $z$ is given by
\ba
\label{eq:rz}
&&r(z)=cH_0^{-1}\, |\Omega_k|^{-1/2} {\rm sinn}[|\Omega_k|^{1/2}\, \Gamma(z)],\\
&&\Gamma(z)=\int_0^z\frac{dz'}{E(z')}, \hskip 1cm E(z)=H(z)/H_0 \nonumber
\ea
where $\Omega_k=-k/H_0^2$ with $k$ denoting the curvature constant, 
and ${\rm sinn}(x)=\sin(x)$, $x$, $\sinh(x)$ for 
$\Omega_k<0$, $\Omega_k=0$, and $\Omega_k>0$ respectively, and
\be
E(z)=\left[\Omega_m (1+z)^3+\Omega_{\rm rad}(1+z)^4+\Omega_k(1+z)^2+
\Omega_X X(z)\right]^{1/2}
\ee
with $\Omega_X=1-\Omega_m-\Omega_{\rm rad}-\Omega_k$, and the dark energy density
function $X(z) \equiv \rho_X(z)/\rho_X(0)$.

CMB data give us the comoving distance to the recombination surface 
$r(z_{CMB})$ with $z_{CMB}=1089$, and the comoving sound horizon 
at recombination\cite{EisenHu98,Page03}
\ba
\label{eq:rs}
r_s(z_{CMB}) &=& \int_0^{t_{CMB}} \frac{c_s\, dt}{a}
=cH_0^{-1}\int_{z_{CMB}}^{\infty} dz\,
\frac{c_s}{E(z)}, \nonumber\\
&=& cH_0^{-1} \int_0^{a_{CMB}} 
\frac{da}{\sqrt{ 3(1+ \overline{R_b}\,a)\, a^4 E^2(z)}},
\ea
where $a$ is the cosmic scale factor, 
$a_{CMB} =1/(1+z_{CMB})$, and
$a^4 E^2(z)=\Omega_m (a+a_{\rm eq})+\Omega_k a^2 +\Omega_X X(z) a^4$,
with $a_{\rm eq}=\Omega_{\rm rad}/\Omega_m=1/(1+z_{\rm eq})$, and
$z_{\rm eq}=2.5\times 10^4 \Omega_m h^2 (T_{CMB}/2.7\,{\rm K})^{-4}$.
The sound speed is $c_s=1/\sqrt{3(1+\overline{R_b}\,a)}$,
with $\overline{R_b}\,a=3\rho_b/(4\rho_\gamma)$,
$\overline{R_b}=31500\Omega_bh^2(T_{CMB}/2.7\,{\rm K})^{-4}$.
COBE four year data give $T_{CMB}=2.728\,$K \cite{Fixsen96}.
The angular scale of the sound horizon at recombination is
defined as $l_a=\pi r(z_{CMB})/r_s(z_{CMB})$ \cite{Page03}.

\begin{figure} 
\psfig{file=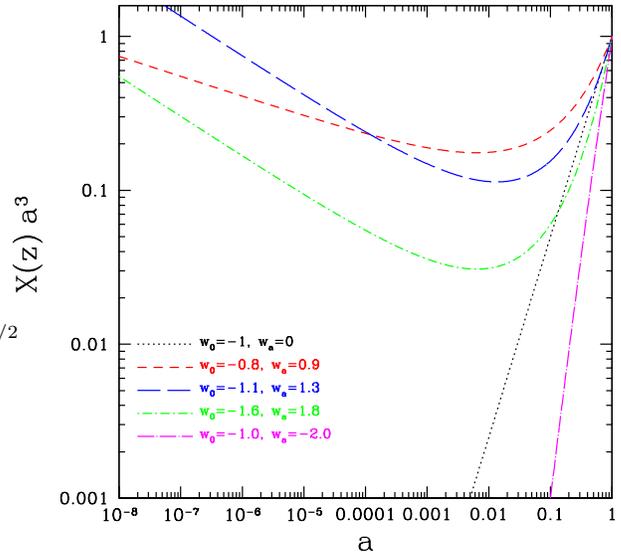,width=3.6in}
\vskip-1cm
\caption[2]{\label{fig:xza3}\footnotesize%
Ratio of the dark energy density $X(z)\equiv \rho_X(z)/\rho_X(0)$
and the matter density $\rho_m(z)/\rho_m(0)=(1+z)^3$ 
for dark energy models
with dark energy equation of state $w_X(z)=w_0+w_a(1-a)$. 
}
\end{figure}

Note that it is important to use the full expression given
in Eq.(\ref{eq:rs}) in making predictions for $l_a$ for
dynamical dark energy models. Fig.{\ref{fig:xza3}}
shows how the dark energy density $X(z)\equiv \rho_X(z)/\rho_X(0)$
compares with the matter density $\rho_m(z)/\rho_m(0)=(1+z)^3$ 
for a two parameter dark energy model
with dark energy equation of state $w_X(z)=w_0+w_a(1-a)$ \cite{Chev01}
which corresponds to $X(z)= a^{-3(1+w_0+w_a)}e^{3w_a(a-1)}$.
For models with $w_0+w_a>0$, the dark energy contribution to
the expansion rate of the universe dominates over that of matter
at high $z$. 
For models that allow significant early dark energy 
(as in the $w_X(z)=w_0+w_a(1-a)$ model), $l_a$
can be underestimated by $20-40$\% if the dark energy
contribution to $r_s(z_{CMB})$ is ignored.\footnote{The
importance of including the dark energy contribution to $l_a$
is also pointed out by \cite{Wright07}.} 

We will show that the CMB shift parameters
\be
R \equiv \sqrt{\Omega_m H_0^2} \,r(z_{CMB}), \hskip 0.1in
l_a \equiv \pi r(z_{CMB})/r_s(z_{CMB}),
\ee
together with $\Omega_b h^2$, provide an efficient summary
of CMB data as far as dark energy constraints go 
(see Sec.IIIA).

SN Ia data give the luminosity distance as a function of redshift,
$d_L(z)=(1+z)\, r(z)$.
We use 182 SNe Ia from the HST/GOODS program \cite{Riess07} and the first 
year SNLS \cite{Astier05}, together with nearby SN Ia data,
as compiled by \cite{Riess07}.
We do not include the ESSENCE data \cite{Wood07}, as these are not yet derived using
the same method as thosed used in \cite{Riess07}.
Combining SN Ia data derived using different analysis techniques 
leads to systematic effects in the estimated SN distance moduli
\cite{Wang00,Wood07}.
Appendix A describes in detail how we use SN Ia data (flux-averaged and
marginalized over $H_0$) in this paper.

We also use the SDSS baryon acoustic oscillation (BAO)
scale measurement by adding the following term to the
$\chi^2$ of a model:
\be
\chi^2_{BAO}=\left[\frac{(A-A_{BAO})}{\sigma_A}\right]^2,
\label{eq:chi2bao}
\ee
where $A$ is defined as
\be
\label{eq:A}
A = \left[ r^2(z_{BAO})\, \frac{cz_{BAO}}{H(z_{BAO})} \right]^{1/3} \, 
\frac{\left(\Omega_m H_0^2\right)^{1/2}} {cz_{BAO} },
\ee
and $A_{BAO}=0.469\,(n_S/0.98)^{-0.35}$,
$\sigma_A= 0.017$, and $z_{BAO}=0.35$
(independent of a dark energy model) \cite{Eisen05}. 
We take the scalar spectral index $n_S=0.95$ as measured by WMAP3
\citep{Spergel06}.\footnote{Note that the \cite{Eisen05} 
constraint on $A$ depends on the 
scalar spectral index $n_S$. Since the error on $n_S$ from WMAP data does not 
increase the effective error on $A$, and the correlation of $n_S$ with
$R$ and $l_a$ is weak, we have ignored the 
very weak correlation
of $A$ with $R$ and $l_a$ in our likelihood analysis.
We have derived $R$ and $l_a$ from WMAP data marginalized over 
all relevant parameters.}

For Gaussian distributed measurements, the likelihood function
$L\propto e^{-\chi^2/2}$, with 
\be
\chi^2=\chi^2_{CMB}+\chi^2_{SNe}+\chi^2_{BAO},
\label{eq:chi2}
\ee
where $\chi^2_{CMB}$ is given in Eq.({\ref{eq:chi2CMB}}) in Sec.IIIA,
$\chi^2_{SNe}$ is given in Eq.({\ref{eq:chi2sn}}) in Appendix A, and
$\chi^2_{BAO}$ is given in Eq.({\ref{eq:chi2bao}}).

We derive constraints on the dark energy density function
$X(z) \equiv\rho_X(z)/\rho_X(0)$
as a free function at $z\leq z_{cut}$, with its value at redshifts
$z_i=z_{cut} (i/n)$ (i=1, 2, ..., $n$), $X(z_i)$, treated 
as $n$ independent parameters estimated from data.
We use $n=3$ and $z_{\rm cut}=1.4$ in this paper.
We use cubic spline interpolation to obtain values of $X(z)$ at other
values of $z$ at $z<z_{cut}$ \citep{WangTegmark04}. 
The number of currently published SNe Ia is very few 
beyond $z_{\rm cut}=1.4$.
For $z>z_{\rm cut}$, we assume $X(z)$ to be 
matched on to either a powerlaw \cite{WangTegmark04}:
\be
X(z)=X(z_{\rm cut}) \left( \frac{1+z}{1+z_{\rm cut}}\right)^{\alpha},
\label{eq:alpowerlaw}
\ee
or an exponential function:
\be
X(z)=X(z_{\rm cut})\, e^{\alpha(z-zcut)}.
\label{eq:alexp}
\ee
We impose a prior of $\alpha \geq -3$ as $\alpha$ is not bounded from below.
Our approach effectively decouples late time dark energy (which is responsible
for the observed recent cosmic acceleration and is probed directly
by SN Ia data) and early time dark energy (which is poorly constrained)
by parametrizing the latter with an additional parameter 
estimated from data.

For comparison with the results of others,
we also derive constraints for models
with dark energy equation of state $w_X(z)=w_0+w_a(1-a)$.
This parametrization has the advantage of not requiring a cutoff
to obtain a finite dark energy equation of state at high $z$ 
(which is not true for the $w_X(z)=w_0+w^{\prime} z $ parametrization),
but it does allow significant early dark energy (which can cause
problems for Big Bang Nucleosynthesis \cite{Steigman06} \footnote{The 
current BBN constraints, $S=0.942\pm 0.030$
($N_{\nu}=2.30^{+0.35}_{-0.34}$) rule out the standard model 
of particle physics ($S=1$, $N_{\nu}=3$) at 1$\sigma$ \cite{Steigman06}.
Given the uncertainties involved in deriving the BBN constraints,
we relax the standard deviation of $S$ by a factor of two, so that the 
standard model of particle physics is allowed at 1$\sigma$.
We find that the resultant BBN constraints do not have measurable effect 
on our dark energy constraints.}
and cosmic structure formation \cite{Sandvik}),
unless a cutoff is imposed. This dilemma illustrates the limited
usefulness of simple parametrizations of dark energy.

For all the dark energy constraints from combining the different
data sets presented in this paper, 
we marginalize the SN Ia data over $H_0$ in
flux-averaging statistics (described in the next subsection), and 
impose a prior of $H_0=72\pm 8\,$(km/s)Mpc$^{-1}$
from the HST Cepheid variable star observations \cite{HST_H0}.

We run a Monte Carlo Markov Chain (MCMC) based on the MCMC engine 
of \cite{Lewis02} to obtain ${\cal O}$($10^6$) samples for each set of 
results presented in this paper. 
For the full CMB analysis we used the WMAP three year temperature and 
polarization \footnote{The main contribution of CMB polarization data is 
the determination of the reionization optical depth.}
power spectra \cite{Spergel06} with version 2 of their 
likelihood code \cite{url} together with theoretical power spectra 
generated by CAMB (with perturbations in dark energy) \cite{Lewis00};
the parameters used are ($\Omega_k$, $\Omega_b h^2$, $\Omega_m h^2$, 
$h$, $A_s$, $\tau$, $n_s$, $\mbox{\bf p}_{DE}$).
For the combined data analysis using CMB shift parameters,
the parameters used are ($\Omega_k$, $\Omega_m$, $h$, $\Omega_b h^2$, 
$\mbox{\bf p}_{DE}$). The dark energy parameter set 
$\mbox{\bf p}_{DE}=w$ for a constant $w_X(z)$,  
$\mbox{\bf p}_{DE}=(w_0,w_a)$ for $w_X(z)=w_0+w_a(1-a)$,
and $\mbox{\bf p}_{DE}=(X(z_1),X(z_2),X(z_3),\alpha)$ for the
general case.
We assumed flat priors for all the parameters, and allowed ranges 
of the parameters wide enough such that further increasing the allowed 
ranges has no impact on the results (with the exception of constraining
$w$ and $(w_0,w_a)$ using CMB data only where we have to impose fixed allowed
ranges for $w$ and $(w_0,w_a)$ since these are not well constrained).
The chains typically have worst e-values (the
variance(mean)/mean(variance) of 1/2 chains)
much smaller than 0.01, indicating convergence.
The chains are subsequently 
appropriately thinned to ensure independent samples.

\section{Results}

\subsection{A Simple and Efficient Method for Incorporating CMB data}

\subsubsection{A roadmap of our method}

We propose a simple and efficient method for dark energy data analysis,
with $\chi^2 = -2\,\ln L=\chi^2_{CMB}+\chi^2_{SNe}+\chi^2_{BAO}$,
where $\chi^2_{CMB}$ is given by constraints on
$(R, l_a, \Omega_b h^2)$ (see Eq.[{\ref{eq:chi2CMB}}] in Sec.IIIA),
$\chi^2_{SNe}$ is given by SN Ia data flux-averaged and marginalized
over $H_0$ (see Eq.[{\ref{eq:chi2sn}}] in Appendix A), and
$\chi^2_{BAO}$ is given by \cite{Eisen05} (see Eq.[{\ref{eq:chi2bao}}]).
In our method, CMB data are incorporated by using constraints on 
$(R, l_a, \Omega_b h^2)$, {\it instead} of using the full CMB power spectra.
In Sec.{\ref{sec:Rla} below, we will show that $(R, l_a, \Omega_b h^2)$
provide an efficient and intuitive summary of CMB data as far as
dark energy constraints are concerned.

\subsubsection{Justification of our method}
\label{sec:Rla}


We have performed MCMC calculations using {\it only}
the full CMB temperature and polarization
angular power spectra from WMAP three year observations,
without assuming spatial flatness, and without
imposing any priors on $H_0$. These calculations are quite time consuming.
We have used these to derived the results in
Fig.{\ref{fig:R}} and Tables I-II.

Fig.{\ref{fig:R}} shows that allowing nonzero cosmic curvature,
the three-year WMAP data give measurements of ($R$, $l_a$, $\Omega_b h^2$) 
that are independent of the dark energy model.\footnote{$R$ 
and $l_a$ are shifted slightly if 
the running of $n_S$ or/and a nonzero tensor to scalar ratio
are considered, and shifted more notably if 
a nonzero neutrino mass is considered\cite{Elgaroy}.
Current CMB data do not require these additional 
parameters\cite{Spergel06}.}
The measurements of ($R$, $l_a$, $\Omega_b h^2$)
differ slightly in a flat universe
because of the correlation of curvature
with other cosmological parameters when spatial flatness
is not assumed.
Table I gives the parameters for the Gaussian fits to
the probability distribution functions of $(R, l_a, \Omega_b h^2, \Omega_m h^2,
r_s(z_{CMB}), r(z_{CMB}))$ from the three-year WMAP data.\footnote{Note 
that CMB data do {\it not} constrain $H_0$ in models with nonzero
curvature due to parameter degeneracies. For example, the dimensionless 
Hubble constant $h=0.50 \pm 0.14$ for a $\Lambda$CDM model with $\Omega_k\neq 0$.
It is the absolute scales of $r_s(z_{CMB})$ and $r(z_{CMB})$ that are well determined
by the CMB data.} 
These fits are independent of the dark energy model assumed.
The constraints on ($\Omega_m h^2$, $r_s(z_{CMB})$, $r(z_{CMB})$)
are also independent of the assumption about cosmic curvature.

\begin{figure} 
\psfig{file=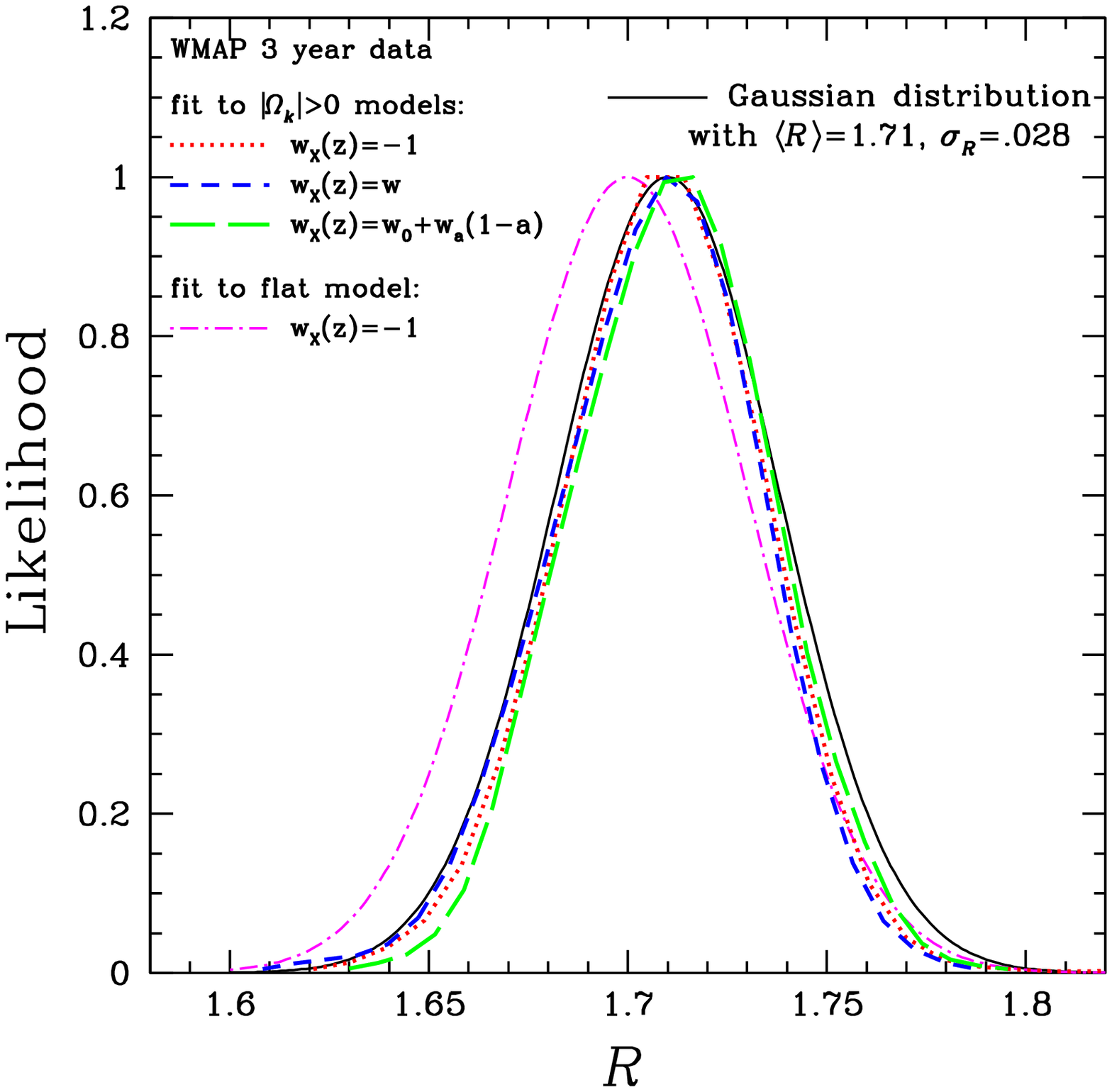,height=2.8in,width=2.8in}\\
\psfig{file=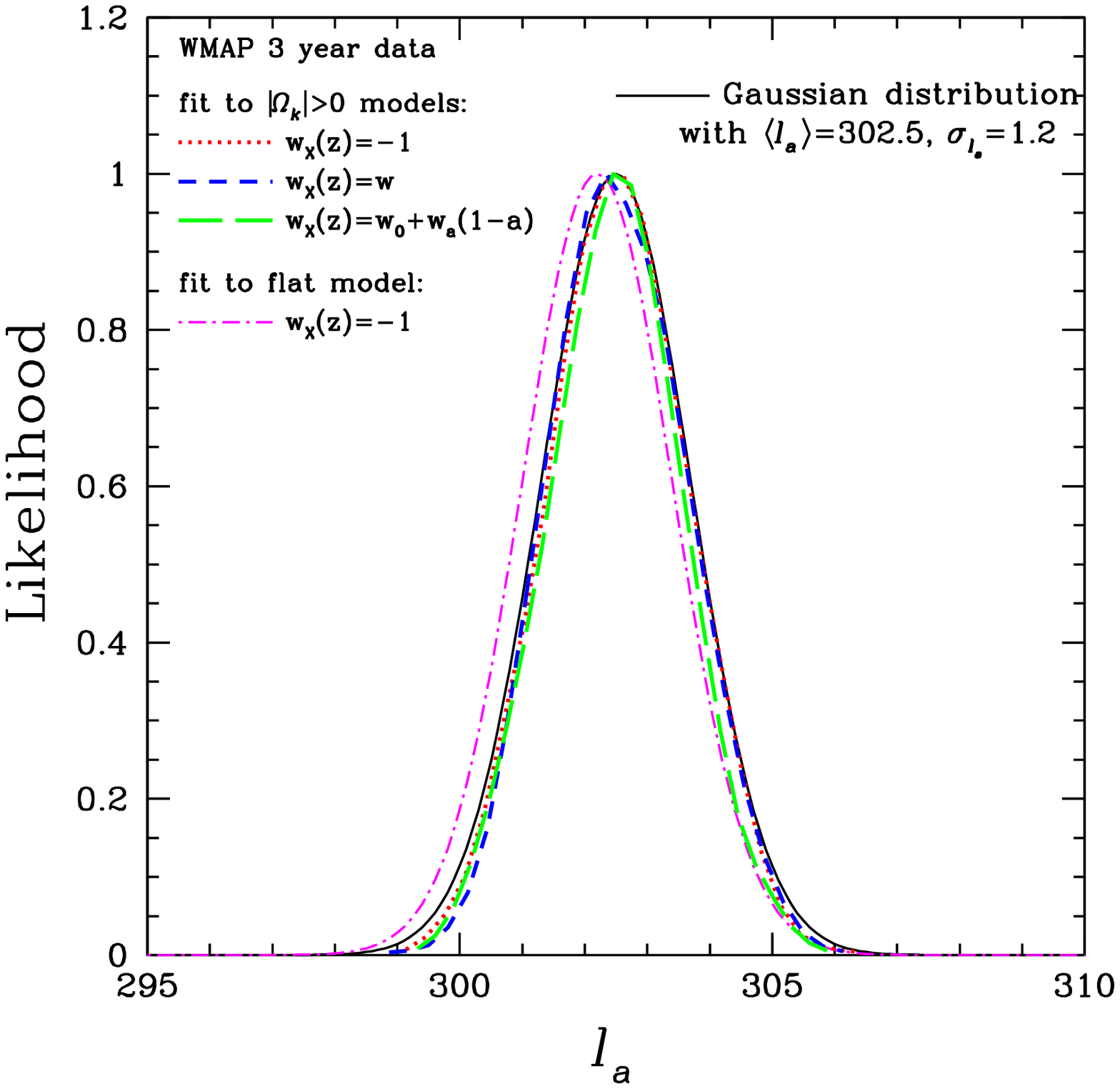,height=2.8in,width=2.8in}\\
\psfig{file=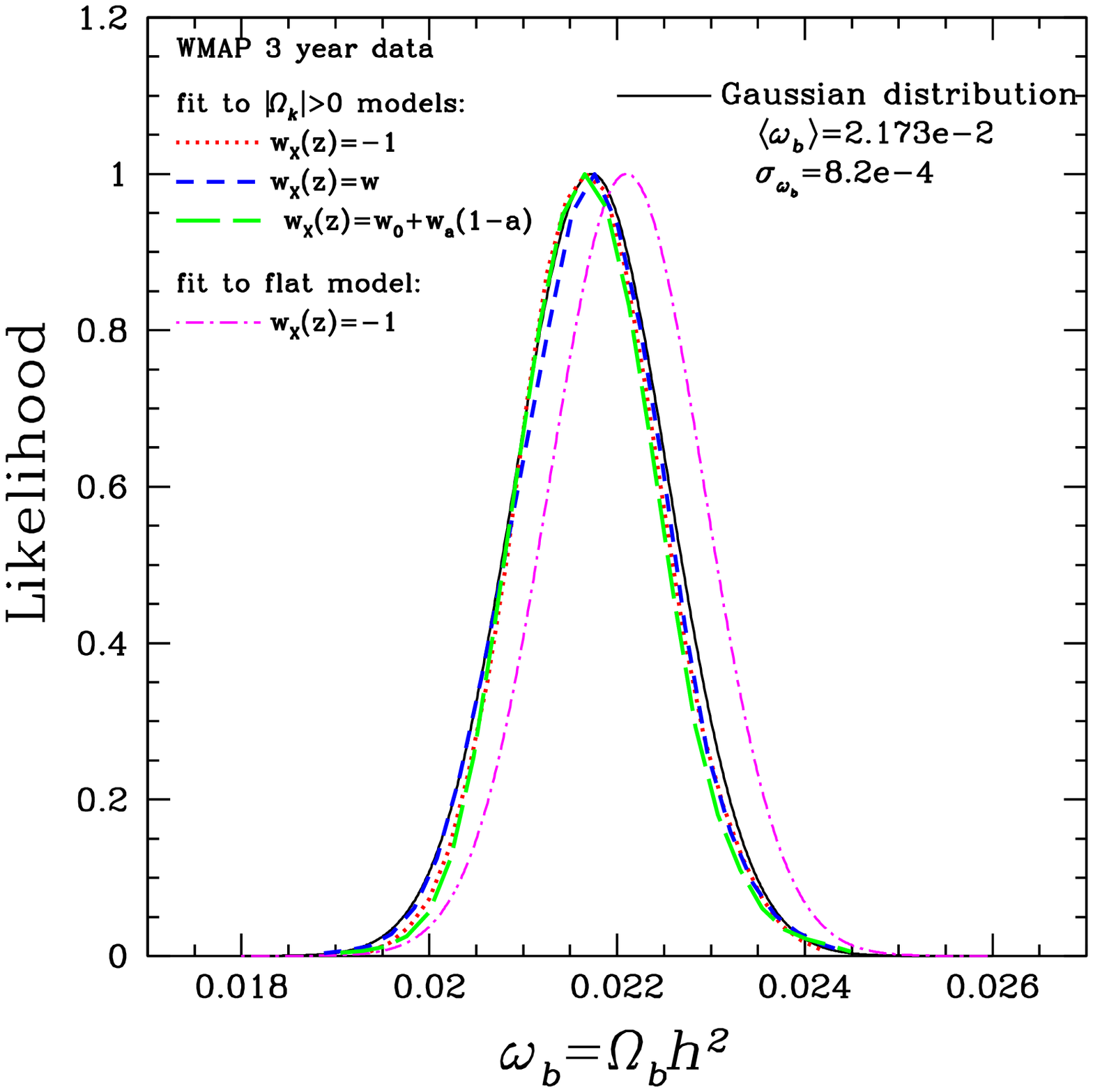,height=2.8in,width=2.8in}
\caption[2]{\label{fig:R}\footnotesize%
The scaled distance to recombination $R$, the angular scale of the sound 
horizon at recombination $l_a$, and the baryon density 
$\Omega_b h^2$ from the three-year WMAP data.
}
\end{figure}

Table II gives the normalized covariance matrices for
$(R, l_a, \Omega_b h^2, \Omega_m h^2, r_s(z_{CMB}), r(z_{CMB}))$
from the three-year WMAP data for a $\Lambda$CDM model
for models with and without curvature. These are appropriate
to use with Table I; models with non-constant dark energy density
give slightly smaller correlations between the parameters.
Note that we have included ($\Omega_m h^2, r_s(z_{CMB}), r(z_{CMB}))$
in Tables I-II to show that although these three parameters
are well constrained by CMB data, they are strongly correlated
with each other\footnote{This high degree of correlation arises 
from how these three parameters are measured. 
The sound horizon at recombination $r_s(z_{CMB})$
is derived primarily using the measurements of $\Omega_b h^2$
and $\Omega_m h^2$ \cite{Page03}, hence is strongly correlated with $\Omega_m h^2$.
The distance to the recombination surface $r(z_{CMB})$ is derived
using $r_s(z_{CMB})$ and the angular scale of the sound horizon
$l_a$ \cite{Page03}, hence is strongly correlated with $r_s(z_{CMB})$
\cite{Page03}}, 
in contrast to the parameters we have chosen,
$(R, l_a, \Omega_b h^2)$.

\begin{table*}[htb]
\caption{The parameters for the Gaussian fits to
the probability distribution functions of $(R, l_a, \Omega_b h^2, \Omega_m h^2,
r_s(z_{CMB}), r(z_{CMB}))$ from the three-year WMAP data,
independent of the dark energy model assumed.}
\begin{center}
\begin{tabular}{lll}
\hline
Parameter & mean & rms variance \\
\hline
\hline
$\Omega_m h^2$ & 0.1284 & 0.0086 \\
$r_s(z_{CMB})$/Mpc & 148.55 & 2.60\\
$r(z_{CMB})$/Mpc & 14305 & 285\\
\hline
\hline
& $\Omega_k\neq 0$&\\
\hline
$R$ & 1.71 & 0.03\\
$l_a$ & 302.5 & 1.2 \\
$\Omega_b h^2$ & 0.02173 & 0.00082 \\
\hline
\hline
& $\Omega_k=0$&\\
\hline
$R$ & 1.70 & 0.03\\
$l_a$ & 302.2 & 1.2 \\
$\Omega_b h^2$ & 0.022 & 0.00082 \\
\hline
 \hline		
\end{tabular}
\end{center}
\end{table*}

\begin{table*}[htb]
\caption{Normalized covariance matrices for 
$(R, l_a, \Omega_b h^2, \Omega_m h^2, r_s(z_{CMB}), r(z_{CMB}))$
fromthe WMAP three year data.}
\begin{center}
\begin{tabular}{lrrrrrr}
\hline
\hline
 & $R$ & $l_a$ & $\Omega_b h^2$ & $\Omega_m h^2$ & $r_s(z_{CMB})$ & $r(z_{CMB})$\\
 \hline
 \hline
& & & $\Omega_k\neq 0$& & &\\
\hline
$R$ &  0.1000E+01 &$-$0.1237E+00  &0.6627E$-$01  &0.9332E+00 &$-$0.8805E+00 &$-$0.8023E+00\\
$l_a$ & $-$0.1237E+00 & 0.1000E+01& $-$0.6722E+00 &$-$0.4458E+00 & 0.5214E+00 & 0.6569E+00\\
$\Omega_b h^2$ &  0.6627E$-$01 &$-$0.6722E+00 & 0.1000E+01 & 0.3731E+00& $-$0.5047E+00& $-$0.5778E+00\\
$\Omega_m h^2$ &  0.9332E+00 &$-$0.4458E+00 & 0.3731E+00 & 0.1000E+01 &$-$0.9882E+00 &$-$0.9605E+00\\
$r_s(z_{CMB})$ & $-$0.8805E+00 &  0.5214E+00 & $-$0.5047E+00 & $-$0.9882E+00 &  0.1000E+01&   0.9859E+00\\
$r(z_{CMB})$ & $-$0.8023E+00 & 0.6569E+00 &$-$0.5778E+00 &$-$0.9605E+00 & 0.9859E+00 & 0.1000E+01\\
\hline
\hline
& & & $\Omega_k= 0$& & &\\
\hline
$R$ &  0.1000E+01&  $-$0.9047E-01&  $-$0.1970E-01&   0.9397E+00&  $-$0.8864E+00 & $-$0.8096E+00\\
$l_a$ & $-$0.9047E$-$01&   0.1000E+01&  $-$0.6283E+00&  $-$0.3992E+00 &  0.4763E+00 &  0.6185E+00\\
$\Omega_b h^2$ & $-$0.1970E$-$01& $-$0.6283E+00& 0.1000E+01 &  0.2741E+00 & $-$0.4173E+00& $-$0.4942E+00\\
$\Omega_m h^2$ & 0.9397E+00& $-$0.3992E+00 & 0.2741E+00 & 0.1000E+01 & $-$0.9876E+00 & $-$0.9594E+00\\
$r_s(z_{CMB})$ & $-$0.8864E+00 & 0.4763E+00 &$-$0.4173E+00 &$-$0.9876E+00 & 0.1000E+01 & 0.9855E+00\\
$r(z_{CMB})$ & $-$0.8096E+00 & 0.6185E+00 &$-$0.4942E+00 &$-$0.9594E+00 & 0.9855E+00 & 0.1000E+01\\
\hline
\hline
\end{tabular}
\end{center}
\end{table*}




We find that there are {\it two} CMB shift parameters, 
$R$ and $l_a$ (with measured values that are nearly uncorrelated, 
see Table II), that are optimal for use in constraining dark 
energy models.\footnote{$R$ has been known as {\it the} CMB shift parameter 
in the past \cite{Bond97,Odman03,WangPia04,WangPia06}.
\cite{Bond97} showed that in an open universe with a 
cosmological constant, there is a degeneracy along the $\delta R=0$ lines,
i.e., models with different values of $\Omega_m$, $\Omega_\Lambda$,
and $h$ that give the same value of $R$ are not distinguishable
except at very low multiples (where cosmic variance dominates),
see Fig.1 of their paper.}
Fig.{\ref{fig:cl2}} shows that both 
$R$ and $l_a$ must be used to describe the complex
degeneracies amongst the cosmological parameters that
determine the CMB angular power spectrum.

Fig.{\ref{fig:cl2}} illustrates the relationship of $R$
and $l_a$ in determining the CMB angular power spectra
for simple models that give the same $R$ or $l_a$ values.
Fig.{\ref{fig:cl2}}(a) shows that models that correspond
to the same value of $R$ but different values of $l_a$
give rise to very different CMB angular power spectra
because $l_a$ determines the acoustic peak structure.
Fig.{\ref{fig:cl2}}(b) shows that models that correspond
to the same value of $l_a$ but different values of $R$
have the same acoustic peak structure in their CMB
angular power spectra, but the overall amplitude of
the acoustic peaks is different in each model because 
of the difference in $R$.\footnote{$R$ is proportional to $\Omega_m h^2$,
which determines the overall height of the acoustic peaks.}
\begin{figure} 
\epsfxsize=\figsize\epsffile{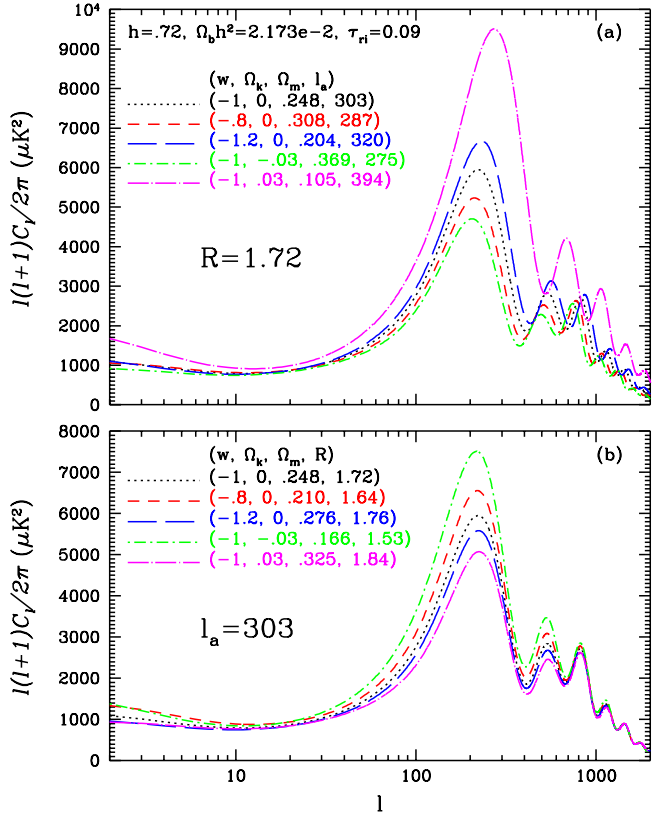}
\caption[2]{\label{fig:cl2}\footnotesize%
CMB angular power spectra for dark energy models
that give the same values of $R$ or $l_a$.
}
\end{figure}

Now we illustrate how using both $R$ and $l_a$ 
helps constrain models with a constant dark energy equation of
state, and zero or small curvature (the class of models
shown in Fig.{\ref{fig:cl2}}).
Fig.{\ref{fig:Rla3}} shows the expected $R$, $l_a$ and
$r_s(z_{CMB})$ as functions of $\Omega_m$ for five models.
For reference, the values for $h$ and $\Omega_b h^2$ have been 
chosen such that the cosmological constant model
satisfies both the $R$ and $l_a$ constraints
from WMAP three year data at the same value of $\Omega_m$
(as in Fig.{\ref{fig:cl2}}).
Note that for the other four models, the $R$ and $l_a$ constraints
cannot be satisfied at the same $\Omega_m$ value.
This is because $R$ and $r_s(z_{CMB})$ have {\it different}
dependences on $\Omega_m$. Models that give the wrong
$R$ and $r_s(z_{CMB})$ values can give the right value
for $l_a$ because $l_a \propto R/r_s(z_{CMB})$.
Using both $R$ and $l_a$ constraints thus helps tighten the
constraint on $\Omega_m$, which leads to tightened constraints
on $w$ or $\Omega_k$.
\begin{figure} 
\epsfxsize=\figsize\epsffile{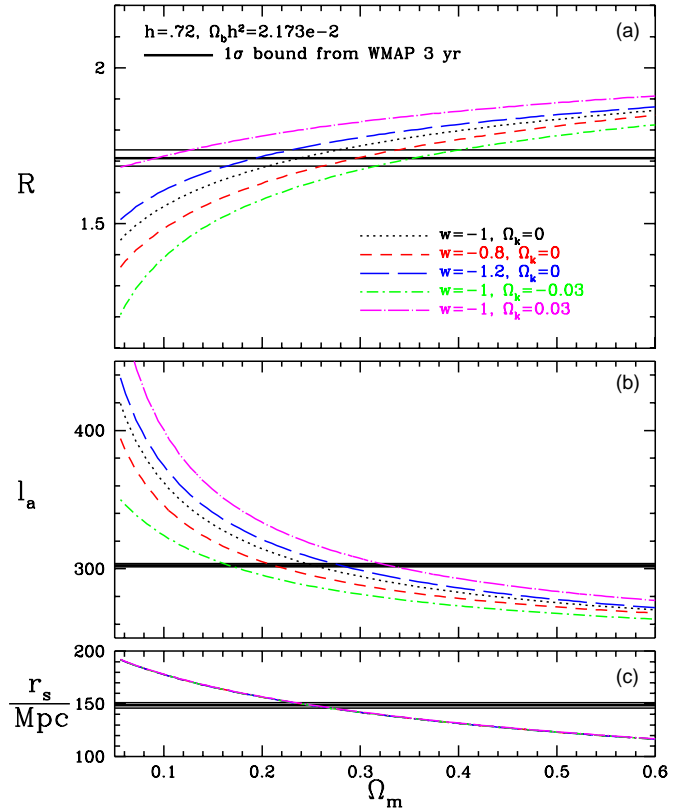}
\caption[1]{\label{fig:Rla3}\footnotesize%
The expected $R$ and $l_a$ as functions of curvature
for five simple dark energy models (with the same line types
as in Fig.{\ref{fig:cl2}}).
}
\end{figure}

When more complicated dark energy models and
nonzero cosmic curvature are considered,
there is a degeneracy between dark energy density 
function $X(z)$ and curvature. The $R$ or $l_a$ 
constraints from CMB can always be satisfied 
with a suitable choice of curvature, but satisfying
the $R$ and the $l_a$ constraints usually
require different values for curvature. 
Thus using both $R$ and $l_a$ constraints
from CMB helps break the degeneracy between
dark energy parameters and curvature.
Fig.{\ref{fig:Rla3w0wa}} demonstrates this by 
showing the expected $R$, $l_a$, and $r_s(z_{CMB})$ as 
functions of curvature for the dark energy models from 
Fig.{\ref{fig:xza3}} (with the same line types).
For reference, the values for $\Omega_m$ and $h$ have been 
chosen such that the cosmological constant model
satisfies both the $R$ and $l_a$ constraints
from WMAP three year data.
Clearly, the $R$ constraint rules out 
closed models with large curvature, while the 
$l_a$ constraint rules out
open models with large curvature.
The vertical dotted lines indicate the 1~$\sigma$ range of $\Omega_k$
from $R$, $l_a$, and $\Omega_b h^2$ constraints from WMAP three year data, 
combined with the data of 182 SNe Ia, and the SDSS BAO measurement.
\begin{figure} 
\epsfxsize=\figsize\epsffile{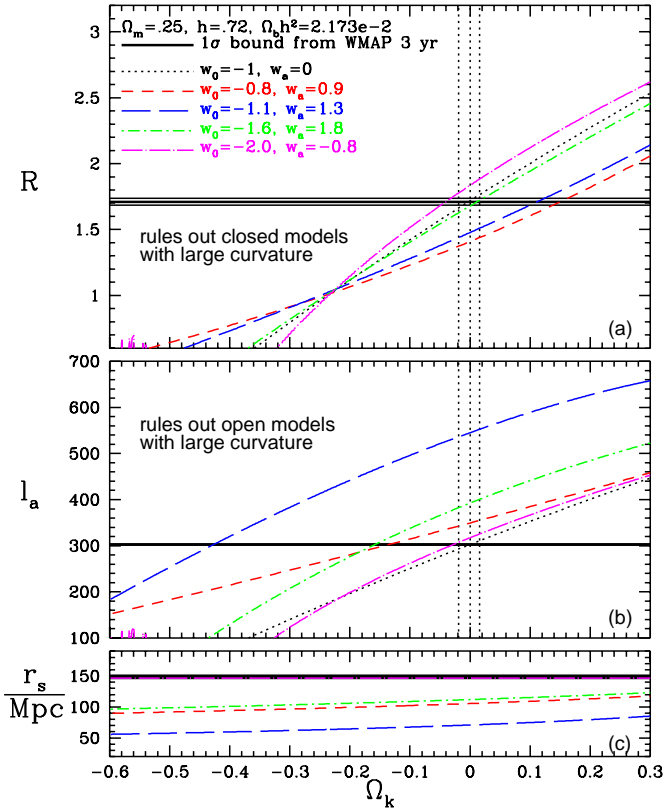}
\caption[1]{\label{fig:Rla3w0wa}\footnotesize%
The expected $R$ and $l_a$ as functions of curvature
for the dark energy models from Fig.{\ref{fig:xza3}}
(with the same line types).
Both $R$ and $l_a$ are needed to
constrain cosmic curvature.
}
\end{figure}

Note that the baryon density $\Omega_b h^2$ should be included 
as an estimated parameter in the data analysis.  
This is because the value of $\Omega_b h^2$ is required
in making a prediction for $l_a$ in a given dark energy model
(see Eq.[{\ref{eq:rs}}]),
and it is correlated with $l_a$ (see Table II).

To summarize, we recommend that the covariance matrix of 
($R$, $l_a$, $\Omega_b h^2$) given in Tables.I-II
be used in the data analysis. To implement this,
simply add the following term to the $\chi^2$ of a given model
with $p_1=R$, $p_2=l_a$, and $p_3=\Omega_b h^2$:
\be
\label{eq:chi2CMB}
\chi^2_{CMB}=\Delta p_i \left[ Cov^{-1}(p_i,p_j)\right]
\Delta p_j,
\hskip .5cm
\Delta p_i= p_i - p_i^{data},
\ee
where $p_i^{data}$ are the mean values given in Table I.
The covariance matrix $Cov(p_i,p_j)$ is obtained by 
multiplying the normalized covariance matrix in Table II 
with $[Var(p_i)\,Var(p_j)]^{1/2}$, with the rms variance
$[Var(p_i)]^{1/2}$ given in Table I.
Note that our constraints on ($R$, $l_a$, $\Omega_b h^2$)
have been marginalized over all other parameters including 
the dark energy parameters.

As a test for the effectiveness of our simple method for
incorporating CMB data, we derived the constraints on 
$w_X(z)=w$ (constant) and $w_X(z)=w_0+w_a(1-a)$
using ($R$, $l_a$, $\Omega_b h^2$), and 
compared with the results from using the full CMB code CAMB.
For both sets of calculations,
we assumed the same flat priors of  $-2\leq w \leq 0$,
$-2\leq w_0 \leq 0$, and $ -6 \leq w_a \leq 3$, 
since $w$ and ($w_0$,$w_a$) are {\it not}
well constrained by using CMB data alone.
The pdf's of $w$ and ($w_0$,$w_a$) span the entire allowed ranges,
and have similar shapes in the two methods.
We did {\it not} assume any priors on $H_0$ since
we want to study CMB data only.
For $w_X(z)=w$ (constant), using ($R$, $l_a$, $\Omega_b h^2$)
gives $w=-0.96\pm  0.57$, while 
the full CMB code CAMB gives $w=-0.97\pm  0.53$.
Using ($R$, $l_a$, $\Omega_b h^2$)
$w_0=-1.0\pm 0.6$ and $w_a=-2.2\pm 2.1$, while
the full CMB code CAMB gives $w_0=-0.9\pm 0.6$ and 
$w_a=-2.4 \pm 2.0$.
These comparisons indicate that our simple method of
incorporating CMB data by using Eq.(\ref{eq:chi2CMB})
is indeed efficient and appropriate as far
as dark energy constraints are concerned.
Since CMB data alone do {\it not} place tight constraints
on dark energy, it is not appropriate to do the comparison
of our method with the full CMB code for dark energy models
with more parameters.

\subsection{Constraints on dark energy}

Because of our ignorance of the nature of dark energy,
it is important to make model-independent constraints
by measuring the dark energy density $\rho_X(z)$ as a free function.
Measuring $\rho_X(z)$ has advantages over measuring dark energy
equation of state $w_X(z)$ as a free function; $\rho_X(z)$ is more
closely related to observables, hence is more tightly 
constrained for the same number of redshift bins 
used \cite{WangGarna,Tegmark02,WangFreese}.
More importantly, 
measuring $w_X(z)$ implicitly assumes that $\rho_X(z)$ does not
change sign in cosmic time (as $\rho_X(z)$ is given by the exponential
of an integral over $1+w_X(z)$); this precludes whole classes of
dark energy models in which $\rho_X(z)$ becomes negative in the future
(``Big Crunch'' models, see \cite{Linde} for an example)\cite{WangTegmark04}.

We have reconstructed the dark energy density function
$X(z)\equiv \rho_X(z)/\rho_X(0)$ by 
measuring its value at $z_i=z_{cut}(i/3)$ (i=1, 2, 3) at $z\leq z_{cut}$,
and parametrized it by either a powerlaw 
($X(z) \propto (1+z)^{\alpha}$) or an exponential function
($X(z) \propto e^{\alpha z}$)
at $z>z_{cut}$ (see Eqs.(\ref{eq:alpowerlaw})-(\ref{eq:alexp})).
We have chosen $z_{cut}=1.4$ as few SNe Ia have been observed
beyond this redshift.
We find that current data allow $\alpha>0$
for $X(z) \propto (1+z)^{\alpha}$ at $z>z_{cut}$,
and require $\alpha<0$ for $X(z) \propto e^{\alpha z}$
at $z>z_{cut}$. This means that assuming powerlaw dark energy
at early times allows significant amount of dark energy
at $z\gg 1$, while assuming exponential dark energy at
early times is equivalent to postulating dark energy
that disappears at $z\gg 1$. The latter is more
physically sensible since dark energy is introduced to
explain late time cosmic acceleration.
Introducing dark energy that is important
at early times could cause problems with Big Bang Nucleosynthesis 
\cite{Steigman06} and formation of cosmic large scale structure 
\cite{Sandvik}.

Fig.{\ref{fig:rhoxz}} shows the reconstructed dark energy density
function $X(z)$ using ($R$, $l_a$, $\Omega_b h^2$)
from the three-year WMAP data, together with 182 SNe Ia and SDSS 
BAO measurement.
The apparent shrinking of the error
contours at $z>z_{cut}$ is due to the use of one parameter to
describe $X(z)$ at $z>z_{cut}$. Future theoretical
work and better data will allow better-motivated description 
of dark energy at early times.\footnote{See for example, \cite{shaf06},
which assumed a flat universe.}
Fig.{\ref{fig:Hz}} shows the corresponding constraints on the cosmic 
expansion history $H(z)$.
\begin{figure} 
\epsfxsize=\figsize\epsffile{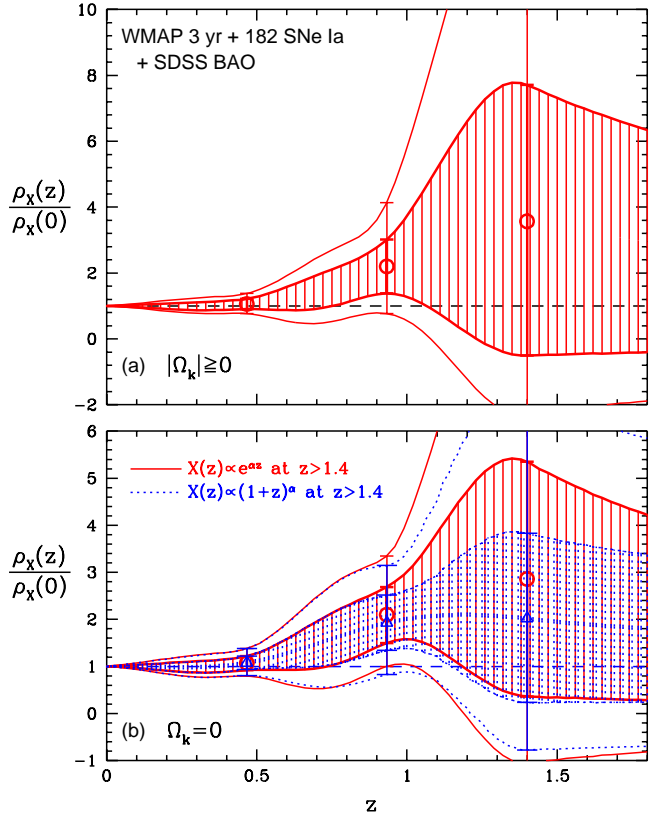}
\caption[1]{\label{fig:rhoxz}\footnotesize%
Constraints on dark energy density using $R$, $l_a$, and $\Omega_b h^2$ 
from the three-year WMAP data,
together with 182 SNe Ia and SDSS BAO measurement.
The shaded areas indicate the 68\% confidence regions, while
the outside contours bound the 95\% confidence regions.
}
\end{figure}
\begin{figure} 
\epsfxsize=\figsize\epsffile{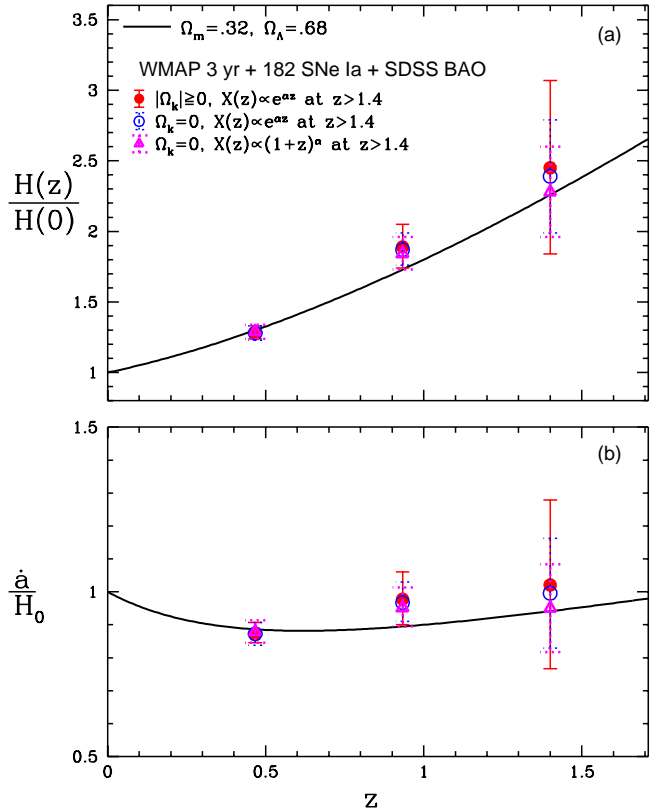}
\caption[1]{\label{fig:Hz}\footnotesize%
Constraints on the expansion history of the universe $H(z)$ 
that corresponds to Fig.{\ref{fig:rhoxz}},
using $R$, $l_a$, and $\Omega_b h^2$ from the 
three-year WMAP data, together with 182 SNe Ia and SDSS BAO measurement.
The error bars indicate the 68\% confidence intervals.
}
\end{figure}

For a flat universe, the dark energy constraints at $z\leq 1$
are nearly independent of the early time assumption about dark energy,
while the dark energy constraint at $z \sim z_{cut}$
is more stringent if $X(z) \propto (1+z)^{\alpha}$
at $z>z_{cut}$. This is as expected.
Because of parameter correlations, stronger assumption about
early time dark energy (the powerlaw form) leads to more stringent
dark energy constraint at late times around $z\sim z_{cut}$.

Without assuming a flat universe,
in the $X(z) \propto (1+z)^{\alpha}$ at $z>z_{cut}$
case, there is a strong degeneracy between curvature and the
powerlaw index $\alpha$. This is as expected since
the curvature contribution to the total matter-energy density 
is also a powerlaw, $(1+z)^2$. $X(z)$ is not well constrained in this case, and is
not shown in Fig.{\ref{fig:rhoxz}}.
When $X(z) \propto e^{\alpha z}$ is assumed at $z>z_{cut}$, 
there is no degeneracy between the exponential index $\alpha$ and curvature.
$X(z)$ is well constrained in this case (see Fig.{\ref{fig:rhoxz}}).

For comparison with the work by others,
Fig.{\ref{fig:w0wa}} shows the constraints on $(w_0,w_a)$ 
for models with dark energy equation of state $w_X(z)=w_0+w_a(1-a)$,
using $R$, $l_a$, and $\Omega_b h^2$ 
from the three-year WMAP data, together with 182 SNe Ia and SDSS BAO measurement.
These are consistent with the results of \cite{Zhao07,Wright07}.
Note that using $w_X(z)=w_0+w_a(1-a)$ implies extrapolation of dark energy 
to early times, which leads to artificially
strong constraints (compared to model-independent constraints)
on dark energy at both early and late 
times. This was noted by \cite{Riess07} as well.
\begin{figure} 
\psfig{file=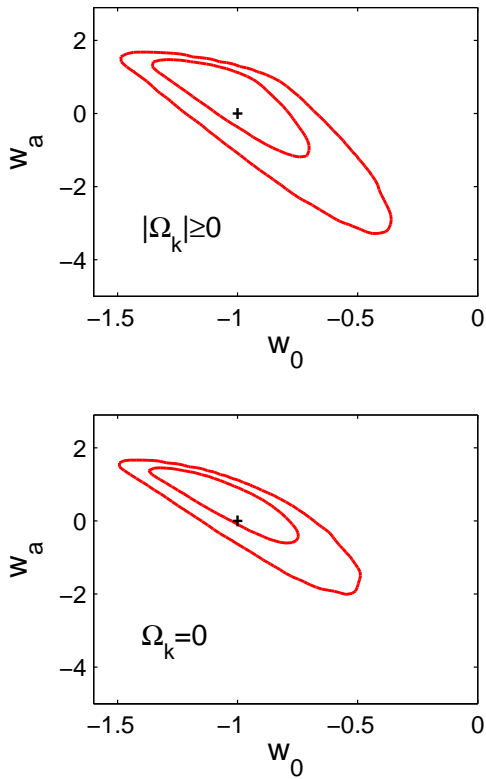,width=6in}
\caption[2]{\label{fig:w0wa}\footnotesize%
Constraints on $(w_0,w_a)$ using $R$, $l_a$, and $\Omega_b h^2$ from 
the three-year WMAP data, together with 182 SNe Ia and SDSS BAO measurement.
The 68\% and 95\% confidence contours are shown.
}
\end{figure}

Comparing Fig.{\ref{fig:rhoxz}}-{\ref{fig:w0wa}} with Figs.3-5 of 
\cite{WangPia06} (for the case of assuming 
$X(z) \propto (1+z)^{\alpha}$ at $z>z_{cut}$), 
it is clear that the constraints on dark energy
have significantly tightened if a flat universe is assumed.

\subsection{Cosmic curvature and dark energy constraints}

Fig.{\ref{fig:ok}} shows the probability distribution function of 
cosmic curvature for different assumptions about dark energy:
the model-independent dark energy density 
$\rho_X(z)$ reconstructed in the last subsection,
the two parameter dark energy model $w_X(z)=w_0+w_a(1-a)$,
and a constant dark energy equation of state. 
A flat universe is allowed at the 68\% confidence level
in all the cases when curvature is well constrained.
$\Omega_k=-0.006_{-0.012}^{+0.013}$$_{-0.025}^{+0.025}$
for assuming that $w_X(z)$ is constant, and
$\Omega_k=-0.002_{-0.018}^{+0.018}$$_{-0.032}^{+0.041}$
for $w_X(z)=w_0+w_a(1-a)$ (68\% and 95\% confidence levels).
Assuming a constant dark energy equation of state
gives the most stringent constraints on cosmic curvature.
The bounds on cosmic curvature are 
less stringent if dark energy density is allowed to be a free
function of redshift,
and are dependent on the assumption
about the early time property of dark energy.
If dark energy is assumed to be an exponential function
at $z>z_{cut}$ ($z_{cut}=1.4$), it is well constrained
by current observational data (see Fig.{\ref{fig:rhoxz}}) and 
negligible at early times. In this case, curvature is
well constrained as well.
If dark energy is assumed to be a powerlaw at early times,
its powerlaw index is strongly degenerate with curvature,
and neither is well constrained.
\begin{figure} 
\psfig{file=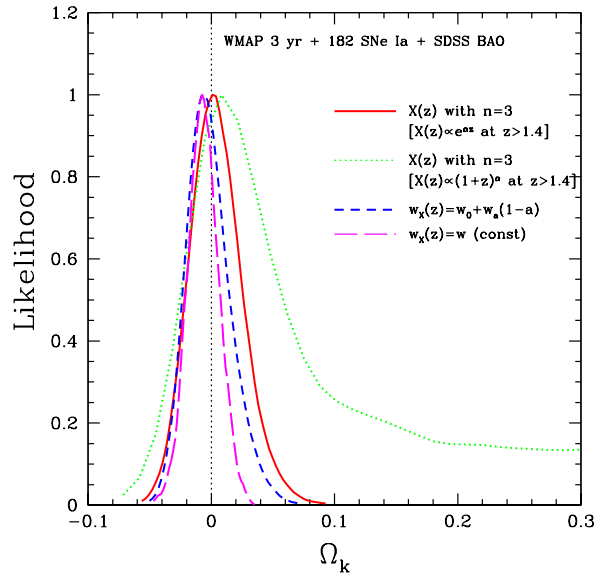,width=3.6in}
\vskip-1cm
\caption[2]{\label{fig:ok}\footnotesize%
The probability distribution function of 
cosmic curvature for different assumptions about dark energy:
the model-independent dark energy density 
$\rho_X(z)$ reconstructed in the last subsection,
the two parameter dark energy model $w_X(z)=w_0+w_a(1-a)$,
and a constant dark energy equation of state. 
}
\end{figure}

\section{Summary and Discussion}

We have presented a simple and effective 
method for incorporating constraints from CMB data into
an analysis of other cosmological data (for example,
SNe Ia and galaxy survey data), when
constraining dark energy without assuming a flat universe.

We find that three-year WMAP data give constraints
on $(R, l_a, \Omega_b h^2, \Omega_m h^2, r_s(z_{CMB}), r(z_{CMB}))$
that are independent of the assumption about dark energy
(see Table I). The constraints on ($\Omega_m h^2, r_s(z_{CMB}), r(z_{CMB}))$)
are also independent of the assumption about cosmic curvature,
but they are strongly correlated with each other and
are not suitable for use in constraining dark energy (see Table II).

We show that there are {\it two} CMB shift parameters,
$R\equiv \sqrt{\Omega_m H_0^2} \,r(z_{CMB})$ 
(the scaled distance to recombination)
and $l_a\equiv \pi r(z_{CMB})/r_s(z_{CMB})$
(the angular scale of the sound horizon at recombination);
these retain the sensitivity to dark energy and curvature of 
$r(z_{CMB})$ and $r_s(z_{CMB})$, and have measured values that are 
nearly uncorrelated with each other (see Table II).
We give the covariance matrix of ($R$, $l_a$, $\Omega_b h^2$)
from the WMAP three year data (see Tables I and II).

We demonstrate that ($R$, $l_a$, $\Omega_b h^2$) provide
an efficient summary of CMB data as far as dark energy constraints
are concerned, and an intuitive way of understanding what the 
CMB does in terms of parameter constraints (see 
Figs.{\ref{fig:cl2}}-{\ref{fig:Rla3w0wa}}).

While completing our paper (based on detailed calculations
that have taken several months), we became aware of Ref.\cite{Elgaroy}.
They also found that using both $R$ and $l_a$ tightens 
dark energy constraints. However, their paper assumed
a flat universe, and used an approximation for $l_a$
that ignores both curvature and dark energy contributions.
We use the exact expression for $l_a$ and derived the covariance
matrix for ($R$, $l_a$, $\Omega_b h^2$) which are based
on the MCMC chains from our full CMB power spectrum calculations
without assuming spatial flatness.

We have used ($R$, $l_a$, $\Omega_b h^2$) from WMAP three year data,
together with 182 SNe Ia (from the HST/GOODS program, the first year 
Supernova Legacy Survey, and nearby SN Ia surveys), 
and SDSS measurement of the baryon acoustic oscillation scale
in deriving constraints on dark energy.
Assuming the HST prior of $H_0=72\pm 8\,$(km/s)Mpc$^{-1}$ \cite{HST_H0},
we find that current observational data 
provide significantly tightened constraints on dark energy models
in a flat universe, and less stringent constraints
on dark energy without assuming spatial flatness 
(see Figs.{\ref{fig:rhoxz}}-{\ref{fig:w0wa}}).
Dark energy density is consistent with a 
constant in cosmic time, with marginal deviations from a 
cosmological constant that may reflect current systematic
uncertainties\footnote{Ref{\cite{Nesseris06}} studied 
the statistical consistency 
of subsets of SNe Ia that comprise the 182 SNe Ia.}
or true evolution in dark energy
(see Figs.{\ref{fig:rhoxz}}-{\ref{fig:Hz}}). 
Our findings are consistent with that of \cite{Riess07}
and \cite{Davis07}.

A flat universe is allowed by current data at the 68\% confidence level.
As expected, the bounds on cosmic curvature are 
less stringent if dark energy density is allowed to be a free
function of cosmic time, and are also dependent on
assumption about dark energy properties at early times 
(see Fig.{\ref{fig:ok}}). The behavior of dark energy at
late times (where it causes cosmic acceleration and is
directly probed by SN Ia data) and at early times (where it
is poorly constrained) should be separated in parameter 
estimation in order to place robust constraints on dark energy
and cosmic curvature (see Sec.IIIB and C).

Future dark energy experiments from both ground and space
\cite{Wang00a,detf,ground,jedi}, together with CMB data from Planck \cite{planck},
will dramatically improve our ability to probe dark energy,
and eventually shed light on the nature of dark energy.

\bigskip

{\bf Acknowledgements}
We thank Jan Michael Kratochvil for being a strong advocate of
marginalizing SN Ia data over $H_0$;
Savas Nesseris, Leandros Perivolaropoulos, and Andrew Liddle
for useful discussions.
We gratefully acknowledge the use of camb and cosmomc.
This work was supported in part by
NSF CAREER grants AST-0094335 (YW).
PM is funded by PPARC (UK).

\appendix

\section{Marginalization over $H_0$ in SN Ia flux statistics}

Because of calibration uncertainties, SN Ia data need to be marginalized
over $H_0$ if SN Ia data are combined with data that are sensitive to
the value of $H_0$. This is the case here (see the next section). We use the angular
scale of the sound horizon at recombination $l_a$ which depends
on $\Omega_m h^2$, while the dimensionless Hubble parameter $E(z)=H(z)/H_0$
(which appears in the derivation of all distance-redshift relations)
depends on $\Omega_m$. Hence a dependence on $H_0$ is implied.
We marginalize the SN Ia data over $H_0$ while
imposing a prior of $H_0=72 \pm 8\,$(km/s)Mpc$^{-1}$ from
HST Cepheid varibale star observations \cite{HST_H0}.

The marginalization of SN Ia data over $H_0$ was derived in \cite{WangGarna}
for the usual magnitude statistics (assuming that the intrinsic dispersion 
in SN Ia peak brightness is Gaussian in magnitudes).
Here we present the formalism for marginalizing SN Ia data over $H_0$ 
in the flux-averaging of SN Ia data using flux statistics (see Eq.[{\ref{eq:chi2flux}}]).
The public software for implementing SN Ia flux averaging with
marginalization over $H_0$ (compatible with cosmomc) is available at
http://www.nhn.ou.edu/$\sim$wang/SNcode/.

Flux-averaging of SN Ia data \cite{Wang00} is needed to minimize the
systematic effect of weak lensing of SNe Ia \cite{lensing}.
\cite{WangPia04} presented a consistent framework for flux-averaging
SN Ia data using flux statistics.
Normally distributed measurement errors are required
if the $\chi^2$ parameter estimate is to be a
maximum likelihood estimator \citep{Press94}.
Hence, if the intrinsic dispersion in SN Ia peak brightness 
is Gaussian in {\it flux}, we have 
\be
\label{eq:chi2flux}
\chi^2_{N_{data}}(\mbox{\bf s})  = \sum_i \frac{ \left[F(z_i) -
F^p(z_i|\mbox{\bf s})\right]^2}{\sigma_{F,i}^2}.
\ee
Since the peak brightness of SNe Ia have been given in magnitudes 
with symmetric error bars, $m_{peak}\pm \sigma_m$, we obtain 
equivalent errors in flux:
\[
\sigma_F \equiv \frac{F(m_{peak}+\sigma_m)-F(m_{peak}-\sigma_m)}{2}.
\]
After flux-averaging, we have
\be
\chi^2 = \sum_i \frac{ \left[\overline{F}(\overline{z}_i) -
F^p(\overline{z}_i|\mbox{\bf s}) \right]^2}{\sigma_{\overline{F},i}^2},
\ee
where $F^p(\overline{z}_i|\mbox{\bf s}_1)=
\left( d_L(\overline{z}_i|\mbox{\bf s}) /\mbox{Mpc} \right)^{-2}$.

The predicted SN Ia flux $F^p(z_i|\mbox{\bf s})
=\left[d_L(z_i|\mbox{\bf s})/{\rm Mpc}\right]^{-2}
\propto h^2$. Assuming that the dimensionless Hubble parameter $h$ is uniformly
distributed in the range [0,1], it is straightforward to integrate over $h$ 
in the probability distribution function to obtain
\be
\label{eq:chi2sn}
p(\mbox{\bf s}|0\leq h \leq 1)= e^{-\chi^2/2}=
\frac{\int_0^1dx \, e^{-g(x)} }{\int_0^1 dx\, e^{-g_0(x)} }
\ee
where
\ba
&&g(x) \equiv \sum_i \frac{ \left[\overline{F}(\overline{z}_i) - x^2
F^p_*(\overline{z}_i|\mbox{\bf s})\right]^2}
{2\sigma_{\overline{F},i}^2}, \nonumber\\
&&g_0(x) \equiv (x^2-1)^2 \sum_i \frac{ \overline{F}(\overline{z}_i)^2}
{2\sigma_{\overline{F},i}^2},
\ea
where $F^p_*(\overline{z}_i|\mbox{\bf s})=F^p(\overline{z}_i|\mbox{\bf s},h=1)$.

\end{document}